\documentclass[10pt]{iopart}
\usepackage{graphicx}
\usepackage{amssymb}
\usepackage{bm}
\usepackage{color}
\usepackage{slashed}
\usepackage{cite}
\usepackage{graphicx}
\usepackage{multicol}
\usepackage{graphicx}
\usepackage{color}
\usepackage{slashed}
\usepackage{CJKutf8}

\begin{document}
\title[Phase diagram of hard core bosons with anisotropic interactions]{Phase diagram of hard core bosons with anisotropic interactions}

\author{Phong. H. Nguyen and Massimo Boninsegni}

\address{
Department of Physics, University of Alberta, Edmonton, Alberta, Canada T6G 2E1}

\ead{m.boninsegni@ualberta.ca}

\begin{abstract}
The phase diagram of lattice hard core bosons with nearest-neighbor interactions allowed to vary independently, from repulsive to attractive, along different crystallographic directions, is studied by Quantum Monte Carlo simulations. We observe a superfluid phase, as well as two crystalline phases at half filling, either checkerboard or striped. Just like in the case of isotropic interactions, no supersolid phase is observed. 
\end{abstract}
%\keywords{Superfluidity, Quantum Monte Carlo simulations, Supersolidity, Quantum Phase Transitions}
\submitto{\CTP}
\maketitle
\ioptwocol
\section{Introduction}\label{intro}

Lattice hard core bosons have been studied theoretically for decades, as a minimal model of a strongly interacting system displaying a superfluid phase at low temperature, as well as a quantum phase transition between superfluid and crystalline ground states. Moreover, significant insight has been gained from the investigation of this model into the possible occurrence of a supersolid phase, and on the role played by point defects such as vacancies and interstitials in its stabilization \cite{matsuda1970,liu1973,batrouni2000,hebert2001,boninsegni2001,heidarian2005,melko2005,wessel2005,boninsegni2005}.
\\ \indent
For a long time, lattice models were regarded as only of academic interest, unable to provide a realistic microscopic description of any actual physical system. However, this state of affairs has drastically changed over the past two decades, as optical lattice technology makes it now feasible, using ultracold atoms and molecules, to create artificial many-body systems accurately realizing the physics contained in these models \cite{jaksch1998,grimm2000,lewenstein2007,windpassinger2013,dutta2015}. Therefore, one may on the one hand, 
compare theoretical predictions and experimental detection  with a degree of accuracy not attainable in experiments with naturally occurring physical system (e.g., solid helium), thereby probing even subtler aspects of the theory; in addition, the possibility of engineering interactions among cold atoms\cite{pethick2008,henkel2010,diaz2013,yu2019,xie2022}  not occurring  in ordinary condensed matter 
(at least not as the dominant interactions) paves the way to the possible observation of novel, exotic phases of matter \cite{boninsegni2012b,schweizer2019,aidelsburger2022}.
\\ \indent
For example,  experimental advances in the spatial confinement and cooling of large assemblies of atoms with finite electric and magnetic dipoles, make it now possible to study many-body systems whose interaction is {\em primarily} dipolar, i.e., anisotropic (attractive or repulsive depending on the relative direction of approach of two particles) and, while not strictly long-ranged, nonetheless extending to distances significantly beyond nearest neighbors, at  experimentally attainable densities. These two features of the pairwise interaction  make it in principle possible to stabilize different crystalline and/or superfluid phases, breaking rotational symmetry. 

The physics of a system of spin zero particles possessing a dipole moment, with all dipole moments aligned along the same direction, has been extensively explored, both theoretically \cite{baranov2008,lahaye2009}  and experimentally (see, for instance, Ref. \cite{henn2014}). In three dimensions, a supersolid phase has been observed in computer simulations of continuous systems \cite{cinti2017,kora2019}. In two dimensions, the situation is more complex; possible supersolid phases arising from speculated ``microemulsion'' scenarios, 
in systems with dipole moments aligned perpendicularly to the plane of particle motion \cite{spivak2004}, have been shown not to occur \cite{moroni2014}. On the other hand, evidence of supersolid phases has been reported in computer simulations of dipolar bosons on the triangular lattice \cite{pollet2010}, whereas on the square lattice no supersolid phases can be stabilized \cite{zhang2015}, even if dipole moments are tilted with respect to the direction perpendicular to the plane of particle motion, a prediction confirmed for continuous systems \cite{cinti2019}\footnote{In three dimensions, as well as lower dimensions if the tilting angle exceeds a critical value, the dipolar interaction becomes purely attractive along specific directions, making the system unstable against collapse. In these cases, the dipolar interaction must be supplemented by a short-range hard core repulsion, whose presence is customarily assumed in standard theoretical studies, to ensure thermodynamic stability.}.
\\ \indent
An interesting theoretical issue that can be addressed, is the relative importance of the long range of the interaction in stabilizing specific, exotic thermodynamic phases, e.g., the supersolid. For example, the phase diagram of lattice dipolar bosons of spin zero, with dipole moments aligned perpendicularly to the plane of motion, is nearly the same as that with only nearest-neighbor interactions, both on the square \cite{hebert2001,zhang2015} and on the triangular lattice \cite{boninsegni2005,pollet2010}. In particular, interstitial supersolid phases exist on the triangular lattice, at particle density 1/3\ (2/3), while they are absent at half filling on the square one. Remarkably, vacancy and interstitial supersolid phases can be stabilized on the square lattice at density 1/4\ (3/4) if {\em only} nearest and next-nearest neighbor interactions are included \cite{chen2008,dang2008}.
\\ \indent
In this paper, we report results of a theoretical investigation of the phase diagram of a system of hard core bosons on the square lattice, interacting via an anisotropic nearest-neighbor potential. We allow the interaction to be repulsive in one direction and attractive in the other. We perform computer simulation of the system using state of the art computational technology, and map out the complete finite temperature phase diagram, which we find to be qualitatively very similar to that of a system of hard core dipolar bosons with aligned dipole moments, tilted with respect to the perpendicular to the plane. The interplay of attraction and repulsion along different directions leads to the occurrence of three phases, a superfluid (SF) one, as well as two crystalline phases at half filling, specifically a checkerboard (CB) and a striped (ST) phase. No supersolid phase is found, either at exactly half filling, consistently with fundamental theoretical arguments \cite{prokofev2005}, or by doping either the CB or ST phase with vacancies or interstitials. A conventional first order quantum phase transition separates both crystalline phases from the superfluid at zero temperature. Basic features of the finite temperature phase diagram are also consistent with what observed in other lattice hard core boson models.
\\ \indent
This paper is organized as follows: in section \ref{model} we describe the model of interest and briefly review the computational technique utilized; in section \ref{res} we present our results, outlining our conclusions in section \ref{concl}.

\section{Model and Methodology} \label{model}
The model of our interest is the well-known lattice hard core Bose Hamiltonian, expressed as follows:
\begin{eqnarray}\nonumber
    \hat H = &-&t \sum_{{\bf r}}\biggl (\hat a^\dagger_{{\bf r}+\hat{\bf x}}\ \hat a_{{\bf r}} + \hat a^\dagger_{{\bf r}+\hat{\bf y}}\ \hat a_{{\bf r}} + h.c.\biggr ) + \\ \nonumber
    &+& \sum_{\bf r} \biggl ( V_{{\bf r},\hat{\bf x}}\ \hat n_{\bf r}\ \hat n_{{\bf r}+\hat{\bf x}} + 
    V_{{\bf r},\hat{\bf y}}\ \hat n_{\bf r}\ \hat n_{{\bf r}+\hat{\bf y}}\biggr ) +\\ 
    &-&\mu\sum_{\bf r}\hat n_{\bf r},
    \label{hamiltonian}
\end{eqnarray}
where the sum runs over all the sites of a square lattice of $N=L\times L$ sites, with periodic boundary conditions. Here, $\hat{\bf x}$ and $\hat{\bf y}$ are the unit vectors in the two crystallographic directions, $\hat a^\dagger_{\bf r}$, $\hat a_{\bf r}$ are the standard Bose creation and annihilation operators, $n_{\bf r}\equiv \hat a^\dagger_{\bf r}\ \hat a_{\bf r}$ is the occupation number for site ${\bf r}$, $t$ is the particle-hopping matrix element and $\mu$ is the chemical potential. The (nearest-neighbor) interaction potential is defined as follows:
\begin{equation}\label{potl}
    V_{{\bf r},\hat{\bf x}} = V, \ \ \ V_{{\bf r},\hat{\bf y}} = \lambda\ V,
\end{equation}
$\lambda$ being a real number. The Hamiltonian (\ref{hamiltonian}) is defined in the subspace of many-particle configurations in which no lattice site can be occupied by more than one particle, which expresses the hard core condition. Thus, the total number of particles $N_P\equiv\sum_{\bf r}\hat a^\dagger_{\bf r}\hat a_{\bf r}$ can take on any integer value from 0 to $N$. The particle density (or, filling) is defined as $\rho\equiv (N_P/N)$, i.e., $0\le \rho\le 1$.

The phase diagram of this model has been extensively investigated in the isotropic $\lambda=1$ case; it is known that, for $-2t < V <2t$, the system displays a single homogeneous fluid phase, undergoing a SF transition at low temperature. On the other hand, if $V > 2t$ a first order quantum phase transition occurs at temperature $T=0$ between a SF and a CB crystal at half filling (i.e., $\rho=1/2$) \cite{hebert2001}. For $V < -2t$, on the other hand, only two  phases occur with $\rho=0$ or 1, phases of intermediate filling (some of them SF) only stabilized by external agents, such as disorder \cite{dang2009}.
\\ \indent
The $\lambda\ne 1$ case, on the other hand, is relatively unexplored, mainly because for a long time a possible experimental realization did not seem possible or straightforward. However, an interaction with that kind of anisotropy can be obtained, in dipolar systems, simply by tilting the direction of alignment of dipole moments with respect to the direction perpendicular to the plane of the lattice (see, for  instance, Ref. \cite{zhang2015} for details). While the dipolar interaction extends significantly beyond nearest neighboring lattice sites, in this work we are interested in assessing which features of the phase diagram arise simply as a result of anisotropy in the short range part of the interaction. Of particular interest is the case $\lambda < 0$, i.e., the interaction is repulsive (attractive) in the $x$ ($y$) direction.

In this work, we systematically investigated the finite temperature phase diagram of Eq. \ref{hamiltonian}, as a function of the three parameters $T$, $\mu$ and $\lambda$,  by means of first principle numerical (Monte Carlo) simulations. We used the worm algorithm in the lattice path-integral representation \cite{prokofev1998}, specifically its implementation described in Ref. \cite{pollet2007}. 
In order to characterize the various phases of the system we computed the superfluid density $\rho_S(T)$ using the well-known winding number estimator \cite{pollock1987}, as well as the static structure factor
\begin{equation}
    S({\bf Q}) = \frac{1}{N^2}\ \biggl \langle \biggl | \sum_{\bf r} \hat n_{\bf r}\ e^{i{\bf Q}\cdot{\bf r}} \biggr|^2\biggr \rangle
\end{equation}
where $\langle...\rangle$ stands for thermal average. Presence of crystalline long-range order is signaled by a finite value of $S({\bf Q})$ for some specific wave vector(s). In particular, ${\bf Q}=(\pi,\pi)$ is the wave vector associated to checkerboard order, while  ${\bf Q}=(\pi,0)$ (and not $(0,\pi)$ because of the way we have defined the interaction, Eq. \ref{potl})  signals stripe order, in both cases at half filling.
Numerical simulations have been carried out on  square lattices of different sizes ($L=32$ being the largest) in order to gauge the importance of finite size effects. 
\section{Results}\label{res}
\begin{figure}
\includegraphics[width=\linewidth]{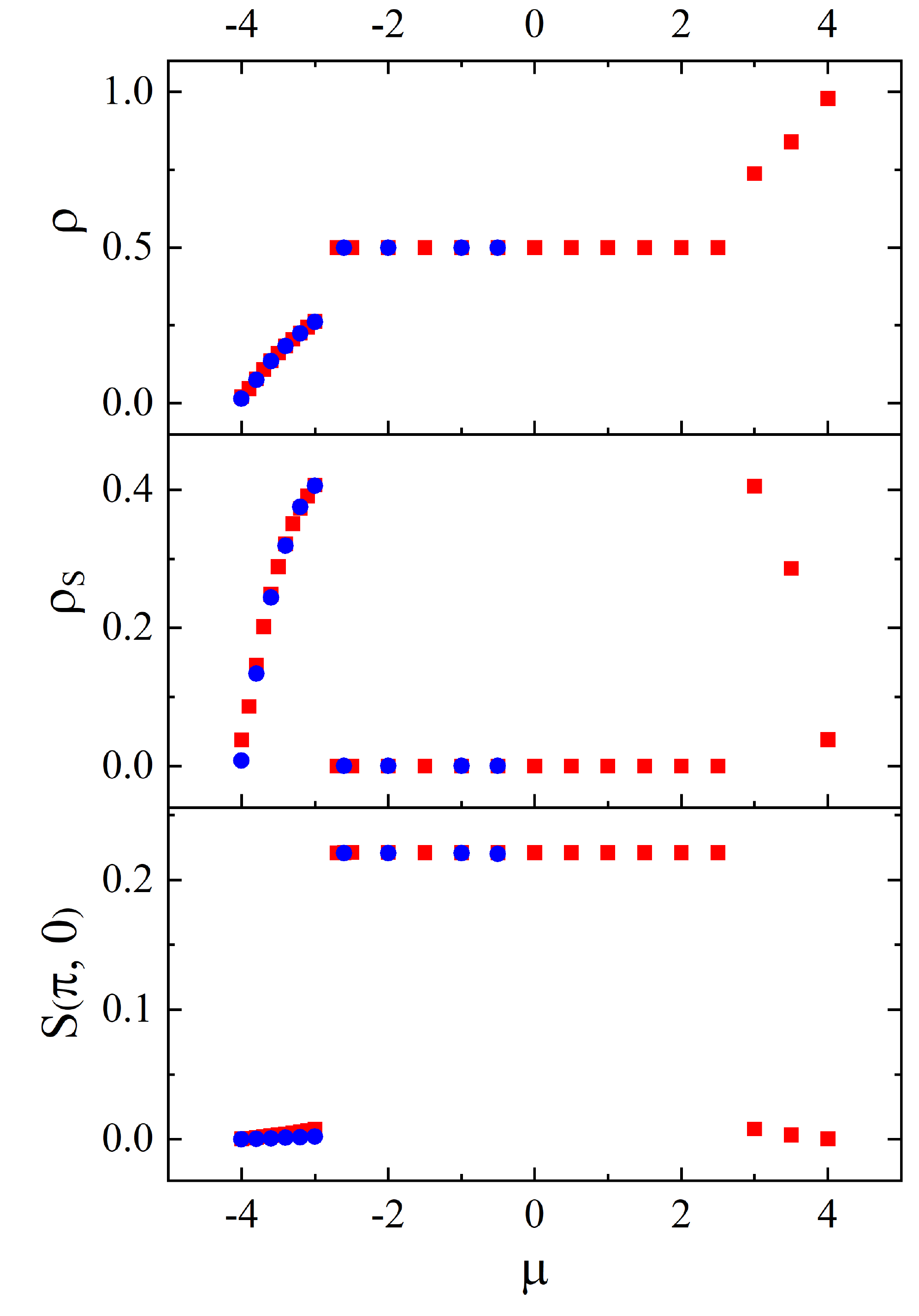}
\caption{Ground state particle density $\rho$ (top panel), superfluid density  $\rho_S$ (middle panel) and static structure factor $S(\pi,0)$ (bottom panel) versus chemical potential $\mu$ (in units of $t$) for $V/t=3$, $\lambda=-1$ and two system sizes $L=8$ (boxes), $L=16$ (circles). Statistical errors are smaller than symbol sizes. }
\label{Fig.1}
\end{figure}
\begin{figure}
\includegraphics[width=\linewidth]{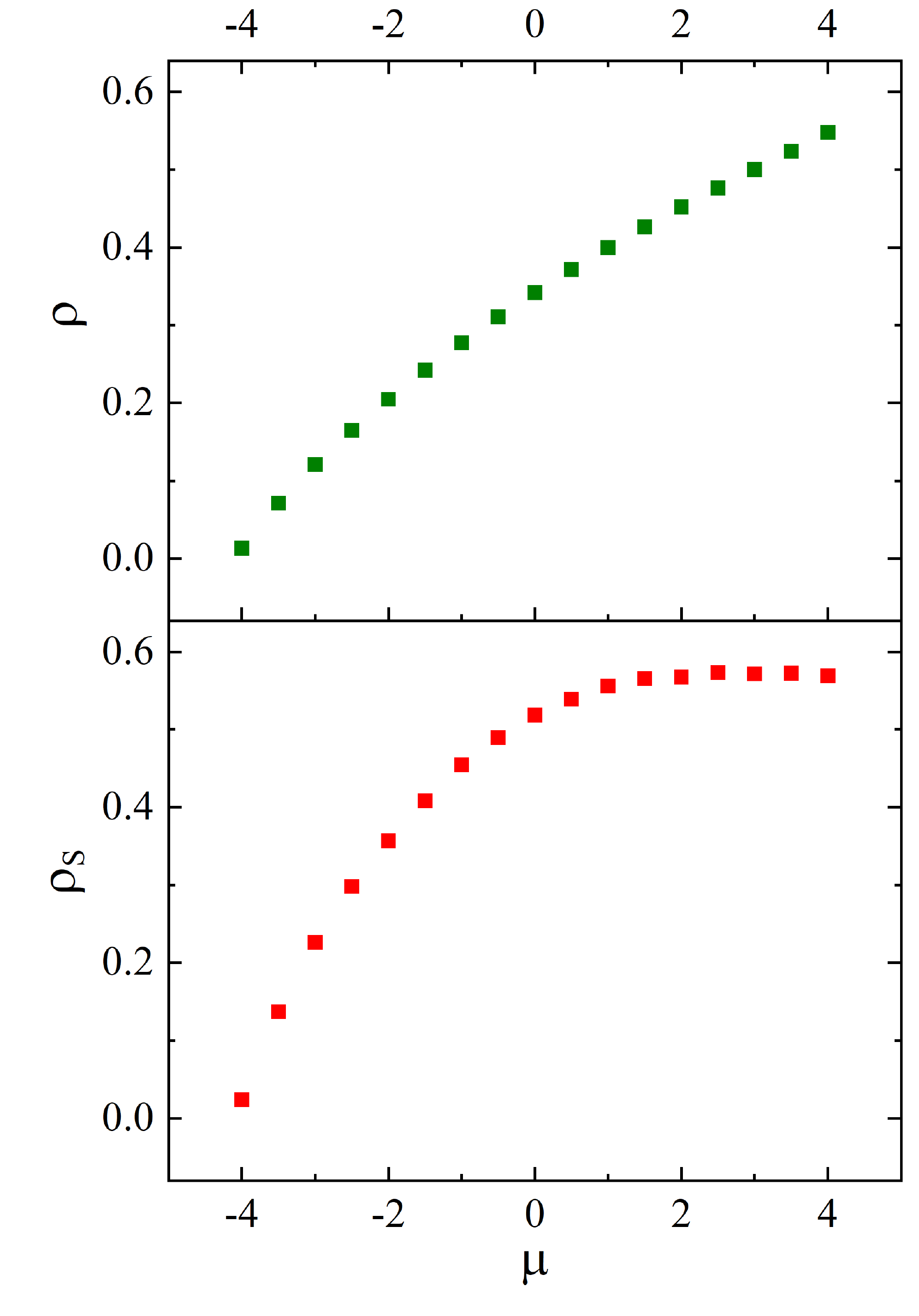}
\caption{Same as Fig. \ref{Fig.1} for $\lambda=0$. }
\label{Fig.2}
\end{figure}

We begin by discussing the physical behavior of the system in its ground state, i.e., at temperature $T=0$. Because we make use of a finite temperature simulation method, this goal is formally accomplished by extrapolating to the $T=0$ limit results obtained at sufficiently low temperatures. In practice, we observe that numerical estimates obtained at temperature $T\sim T_L\equiv (t/L)$ are indistinguishable from the extrapolated ones, within the statistical uncertainties of the calculation, i.e., physical estimates obtained at temperature $T_L$
can be regarded as essentially ground-state estimates. 
\\
\indent
As an example of the procedure followed to characterize the phases of the system, Fig. \ref{Fig.1} illustrates ground state results for the case $\lambda=-1$ and $V=3t$. In this case, the system displays two distinct phases, a superfluid one (possessing no crystalline order) away from half filling, and a non-superfluid crystalline (striped) one at half filling, with a first order phase transition between the two.
\\ \indent
Specifically, the top panel shows the particle density $\rho$ computed as a function of the chemical potential $\mu$ (in units of $t$). The density  jump as $\rho$ approaches half filling signals a first-order phase transition, with coexistence of a superfluid phase (middle panel shows the superfluid density $\rho_S$) and one that features striped crystalline order (bottom panel shows the static structure factor $S(\pi,0))$. As one can see, no supersolid phase exists, i.e.,  in which {\em both} $\rho_S$ and $S(\pi,0)$ are finite. The superfluid density is everywhere finite, except at half filling, whereas the static structure factor is finite only at half filling. This can be contrasted with the physics described in Fig. \ref{Fig.2} for $V/t=3$ and $\lambda=0$. In this case, all curves are everywhere continuous, i.e., no phase transition occurs.
\begin{figure}
\includegraphics[scale=0.32]{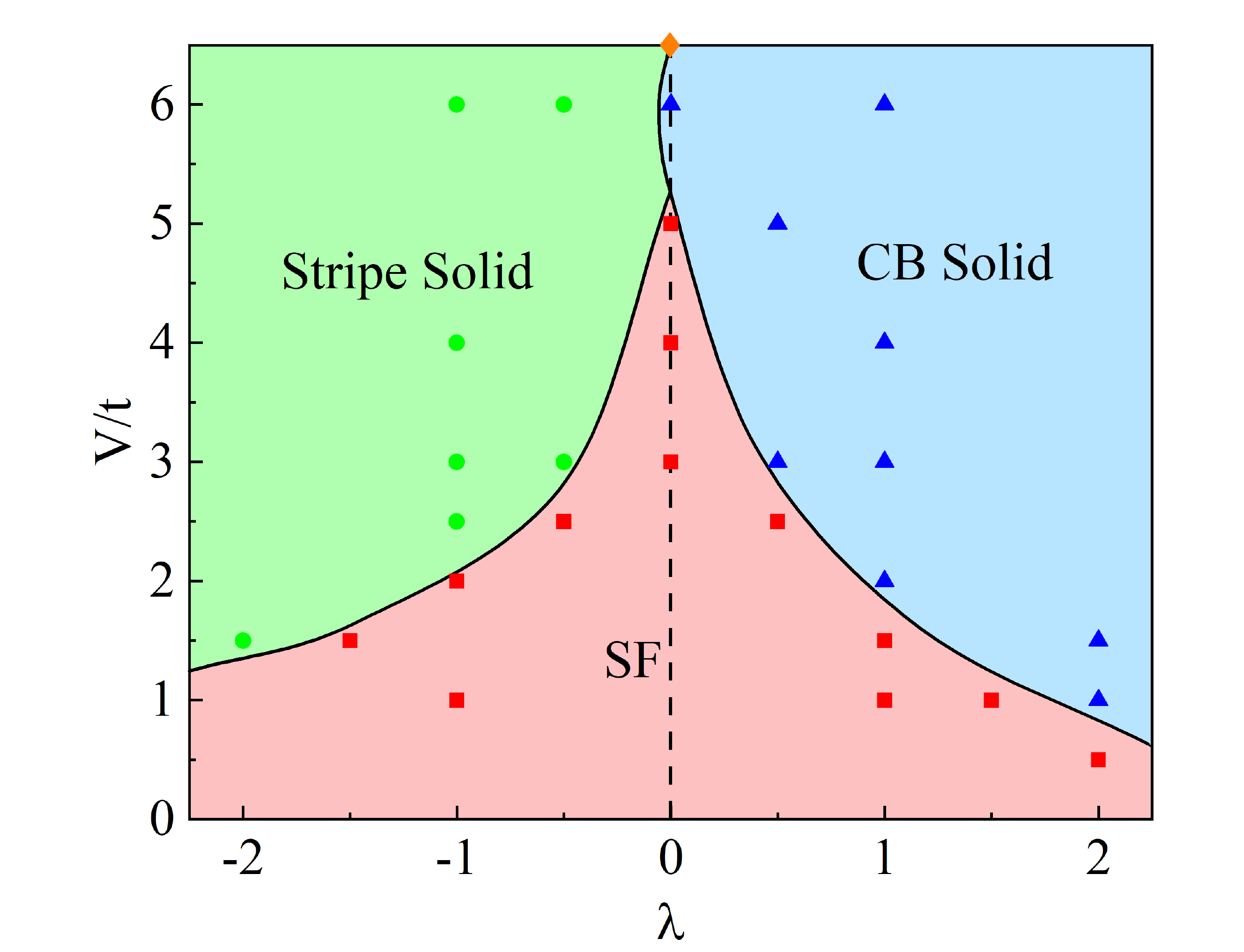}
\caption{Ground state phase diagram at half filling as a function of the interaction strength $V$ and anisotropy parameter $\lambda$. The system features a checkerboard solid (triangles), a stripe solid (circles), and a superfluid phase (boxes). In the $V/t \rightarrow \infty$ limit, the CB solid and the stripe solid lines meet at $\lambda=0$ (diamond at the top). Solid lines separating the various phases indicate first order quantum phase transitions.}
\label{Fig.3}
\end{figure}
\indent
In Fig. \ref{Fig.3}, we present the ground state phase diagram at half filling in the $V - \lambda$ plane, obtained from an analysis of results such as those shown in Figs. \ref{Fig.1} and \ref{Fig.2}, showing the boundaries between the three difference phases that are observed, namely the SF as well as CB and striped  solid phases. Lines separating all the phases indicate first order quantum phase transitions.
This phase diagram is qualitatively very similar to that of a system of tilted dipolar lattice bosons \cite{zhang2015}. As expected, in the weak coupling limit (i.e., at low interaction strength $V$) the system is in the SF phase for any value of the anisotropic parameter $\lambda$. On the other hand, as $V$ is increased to a critical value $V_{c}(\lambda)$ the system transitions into a checkerboard (striped) solid phase when $\lambda>0$ ($\lambda<0$) through a first-order phase transition. As $\lambda\to 0$, superfluid order is resilient for greater values of $V$; for $V \gtrsim 5t$, the system can only be in one of the two crystalline phases. 
\\ \indent
Along the $\lambda=0$ line, the CB phase is favored over the striped one by quantum fluctuations, as one can ascertain through standard second order perturbation theory.
Thus, the line separating the stripe from the CB phase at large values of $V$ is not exactly vertical, as CB order is favored for $\lambda\to0^-$. This is seemingly the most remarkable difference between the anisotropic nearest-neighbor interaction considered in this work, and the tilted, full dipolar interaction \cite{zhang2015}.

\begin{figure}
\includegraphics[width=\linewidth]{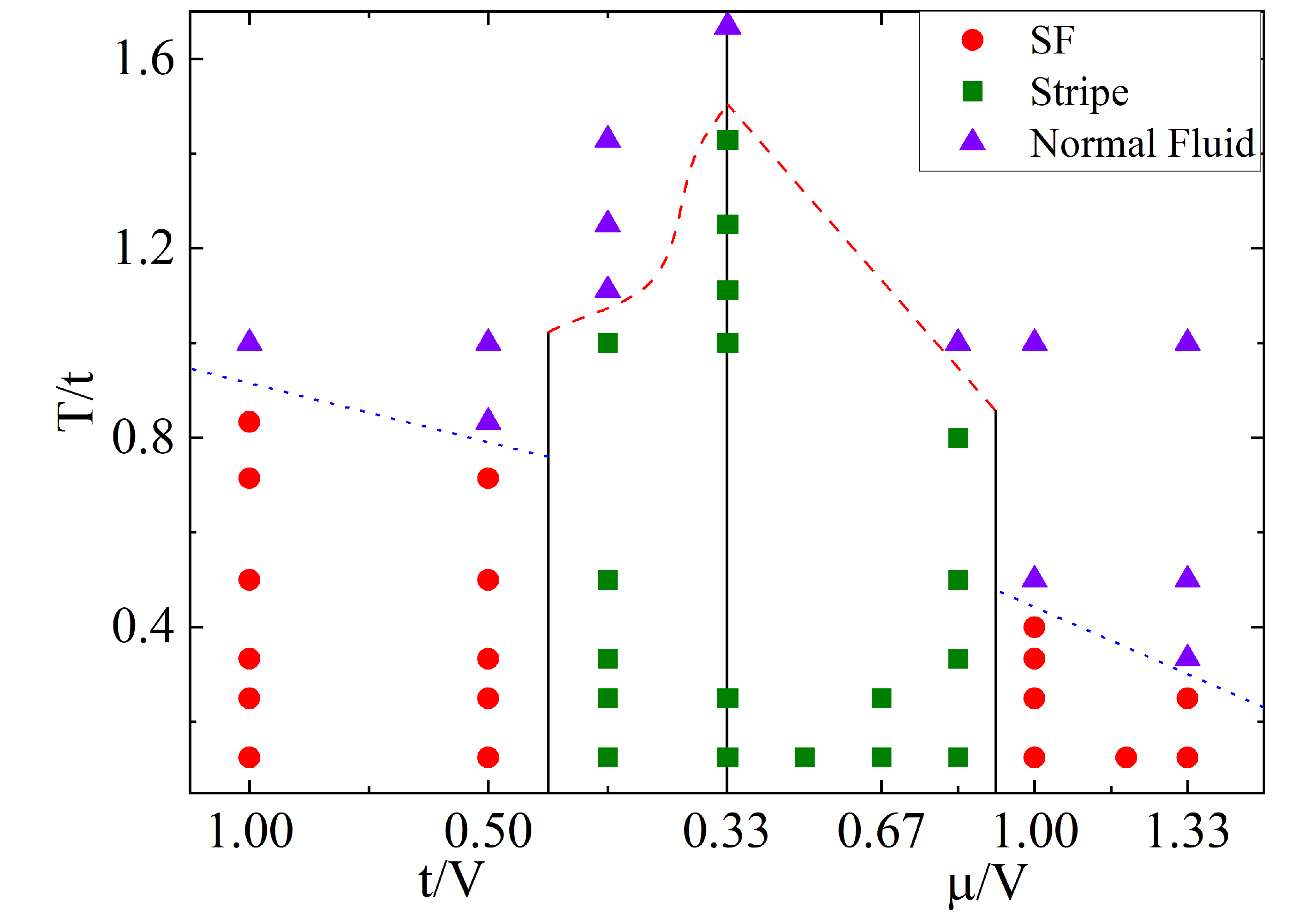}
\caption{Finite temperature phase diagram for $\lambda=-1$ for two representative cuts. Left panel show results for fixed $\mu=t$; the right panel is for fixed $V=3\ t$. Solid lines indicate first-order phase transitions between stripe solid (boxes) and superfluid  (circles) or normal fluid (triangles) phases. Dotted lines show Berezinskii-Kosterlitz-Thouless transitions between normal fluid ( triangles) and superfluid phases. Dashed lines correspond to second-order transitions in the Ising universality class between stripe solid and normal fluid.  }
\label{Fig.4}
\end{figure}

Fig. \ref{Fig.4} shows two representative cuts of the finite-temperature phase diagram of the system in the ($T,\mu,V$) space. We consider for definiteness the case $\lambda=-1$, for which the ground state at half filling is a stripe solid for $V\gtrsim 2.5 \ t$. The left panel has fixed chemical potential ($\mu=t$), while the right panel has fixed interaction strength ($V=3\ t$).

At constant temperature, the striped crystal is separated from the (super)fluid
phase by a first-order (quantum) phase transition, shown by solid lines in Fig. \ref{Fig.4}. On the other hand, at constant $V$ or $\mu$ superfluid order is destroyed by thermal fluctuations in favor of a normal fluid via a Berezinskii-Kosterlitz-Thouless transition, shown by the dotted lines, while the striped crystal melts into a normal fluid via a second-order transition in the Ising universality class (dashed curve in Fig. \ref{Fig.4}). Altogether, this is a fairly conventional phase diagram, qualitatively identical to that in the $\lambda > 0$ sector, with no supersolid phase.

\section {Conclusions} \label{concl}
In this work we investigate the phase diagram of a system of hard core bosons on the square lattice, with a short-ranged (nearest-neighbor) interaction that can vary independently along the two crystallographic directions, to include the case of attraction along one direction and repulsion along the other. This kind of anisotropic interaction can be realized experimentally with a system of dipolar atoms or molecules, confined to a 2D optical lattice, if dipole moments are aligned at a varying (tilting) angle with respect to the direction perpendicular to the plane of particle motion.
\\ \indent
The main goal of this study are {\em a}) to gain understanding of the phases that can arise as a result of the anisotropy of the interaction, {\em b}) to gauge the importance of its extended range, by comparing our results to those obtained in similar studies \cite{zhang2015} in which the full dipolar interaction is considered.
With respect to the case in which the interaction is isotropic, the only difference is the appearance of a different crystalline phase, displaying stripe order. Otherwise, the phase diagram is qualitatively identical to that for isotropic interaction, with a single crystalline phase at half filling, and no supersolid phase. The phase diagram is also qualitatively identical to that observed by allowing the interaction to take on the full dipolar form, i.e., extending beyond nearest neighbors, with varying degree of anisotropy corresponding to different tilting angles \cite{zhang2015}. This is consistent with what observed for purely repulsive, isotropic interactions, where the presence of a $1/r^3$ dipolar tail does not fundamentally alter the phase diagram, e.g., on the triangular lattice \cite{pollet2010}. Altogether, this study reinforces the notion that lattice geometry (i.e., triangular versus square), more than the detailed form of the interaction, is at the root of the occurrence of supersolid phases.
\\ \indent
It has been suggested in a recent study \cite{wu2020}, that in the presence of anisotropic interactions of dipolar, form supersolid phases  may be stabilized, away from half filling, by allowing the interactions to extend to next-nearest neighboring distances. We have not explored this aspect here, but the contention seems plausible and not particularly surprising, given that this was shown to be the case for isotropic repulsive interactions \cite{dang2008} (i.e., the $\lambda > 0$ sector of Eq. \ref{potl}), and the fact that the physics of the system in the positive and negative $\lambda$ sectors is essentially the same, as shown in this study.

%\begin{figure}
%\includegraphics[width=7cm]{Phase_diagram.pdf}
%\caption{$P-\upsilon$ diagram for the HAdS BH and the BAdS BH with $g=1$. The dotted curves correspond to the HAdS BH, the solid curves represent the BAdS BH.}
%\label{fig1}
%\end{figure}

\section*{Acknowledgments}
This work was supported by the Natural Sciences and Engineering Research Council of Canada. Computing support of ComputeCanada is acknowledged.

\section*{References}
\bibliographystyle{iopart-num}
\bibliography{refs.bib}

\end{document}